\documentclass[aps,prl,twocolumn,showpacs,superscriptaddress,groupedaddress]{revtex4} 
\usepackage{dcolumn}   
\usepackage{bm}        
\usepackage{graphicx,amssymb,amsmath}
\usepackage{color}
\usepackage{etoolbox}

\begin{document}

\newlength\figurewidth 
\setlength\figurewidth{7cm}

\title{Using an ion trap with two temperature reservoirs to explore nonequilibrium physics}

\author{Christian Vaca}
\affiliation{Department of Physics \& Astronomy}

\author{Kuang Chen}
\affiliation{Department of Physics \& Astronomy}

\author{Eric Hudson}
\affiliation{Department of Physics \& Astronomy}

\author{Alex J.~Levine}
\affiliation{Department of Physics \& Astronomy} 
\affiliation{Department of Chemistry \& Biochemistry}
\affiliation{Department of Biomathematics\\
UCLA, Los Angeles, CA 90095-1596}

\date{\today}

\begin{abstract}
Version 1.0 Resubmission\\
Using a combination of molecular dynamics simulations and fundamental statistical mechanics, we 
analyze the position and velocity distribution of a trapped ion immersed in two ideal gases at differing temperatures.
Such a system has been realized in the recently developed MOTion trap architecture. This system 
has the potential to serve as platform for studying nonequilibrium statistical mechanics in a controlled environment.
As examples, we demonstrate a non-Maxwell Boltzmann velocity ionic distribution in the trap, the breakdown of the
fluctuation-dissipation theorem, and position-velocity sorting, wherein high-velocity ionic states are over represented at points of 
high potential energy.  We propose experiments to test these predictions.
\end{abstract}


\maketitle

A wide variety of natural processes occur far from equilibrium.  
Their complex phenomenology is, in many cases poorly understood, but has wide-ranging implications from heart dynamics~\cite{Stanley:95}
to climate modeling~\cite{Scheffer:09}.
These systems may exhibit elaborate spatiotemporal patterns, such as in Rayleigh-Benard convection~\cite{Cross:93}, 
oscillatory chemical reactions~\cite{Science-Epstein-01}, or in swarms of active swimmers~\cite{Bartolo:13}.  
Some theorems~\cite{Morris-PRL-93,Jarzynski:93}  have been put forward regarding their (typically large) 
fluctuations~\cite{Derrida:07}.  The number, however, of well-characterized nonequilibrium steady-states is limited. 
To date, soft matter systems such as active gels~\cite{Mizuno:07}, colloids~\cite{Chaikin:13}, or fluids~\cite{Cates-PRE:11, Marchetti-SoftMatter:12} 
have been considered as prototypical models.  Developing even simpler model systems amenable to precise control of their nonequilibrium state is 
essential for further progress in the field.

We investigate such a prototypical nonequilibrium system consisting of a single ion immersed in two non-interacting 
ideal gases, which can be realized in a recently developed hybrid atom-ion trap~\cite{Rellergert:13,reviews1,reviews2,reviews3,reviews4,reviews5}.  
There, the single ion, held in a radio-frequency (rf) trap, interacts with either two-independent laser-cooled buffer gases or one laser-cooled buffer gas and a 
low pressure background gas. Because the neutral-ion collision cross section is orders of magnitudes larger than neutral-neutral collision cross sections, the interaction of the buffer 
gases with one another can be neglected, so that the buffer gas species may have large temperature differences, driving the system out of equilibrium. 
The rf trap also generates micromotion~\cite{Leibfried:03}, further driving the ion out of equilibrium~\cite{DeVoe:09, Chen:13}. The two-temperature buffer gas 
system, which is the focus of the current manuscript, produces a well-defined and highly controllable nonequilibrium steady-state. In experiment both forms of nonequilibrium 
effects may play a role, but, by changing the temperature difference between the two buffer gases one can distinguish their effects.  In our theoretical calculations presented here, 
we can independently examine these two thermodynamic driving terms by comparing results in a model rf trap and a hypothetical static one, which eliminates micromotion 
effects.

We show here that an ion in either trap with a two-temperature buffer gas should exhibit striking 
nonequilibrium features. We focus on three in particular.  We demonstrate (i) large departures from a Maxwellian velocity distribution 
and (ii) the non-factorizability of the joint position--velocity  probability distribution.  
This factorizibility is a hallmark of classical equilibrium statistical mechanics. In the nonequilibrium 
state, the ion exhibits  {\em position-velocity sorting} wherein ionic high velocity states are overpopulated 
in regions of high potential energy relative to regions of low potential energy.  
This property allows one to construct, in principle, a heat engine, even in the static trap,
using the edges and  center of the trap as heat sources and sinks, respectively. We also demonstrate  (iii) the 
breakdown of the fluctuation-dissipation theorem from a comparison of ionic mobility and diffusion. There will be 
interesting nonequilibrium effects in ionic transport.   These three features demonstrate that 
this system is a rich playground for exploring nonequilibrium physics in a precise and controllable manner. 
We conclude by suggesting experiments to observe these features and propose future research directions.

We begin with a one-dimensional model of a trapped ion of mass $M$ interacting with two non-interacting ideal gases of masses 
$m_{\rm h}$ and $m_{\rm c}$ at differing temperatures -- see the inset of Fig.~1 for 
a schematic representation of the ion (green) in a potential interacting with hot (red) and cold (blue) atoms.  
The time-dependent probability distribution $P(x,v,t)$ of the ion velocity $v$ and 
position $x$ obeys the master equation
\begin{widetext}
\begin{equation}
\left(\partial_t+v\, \partial_x+\frac{F(x,t)}{M}\,\partial_v\right)P(x,v,t)=\int dv'\, W(v|v')P(x,v',t)-\int dv'\, W(v'|v)\,P(x,v,t).
\label{eq:MasterEquation}
\end{equation}
\end{widetext}
The left side of Eq.~\ref{eq:MasterEquation} contains the time derivative of the probability density and two streaming terms representing the advection of probability in 
position space due to the ion velocity $v$ and the advection of probability in velocity space due to the acceleration of the ion in response to the trapping force $F(x,t)$.  
The terms on the right side of Eq.~\ref{eq:MasterEquation} represent the effect of collisions between the ions and the buffer gases. 
These collisions generate ion velocity transitions from $v_{1}$ to $v_{2}$ with probability per unit time $W(v_{2}|v_{1})dv_{2}$.
We assume that the two buffer gases are themselves in equilibrium states with Maxwell-Boltzmann (MB) velocity distributions corresponding to 
temperatures $T_{\rm h} > T_{\rm c}$ and number densities $c_{\rm h}$, $c_{\rm c}$ respectively.
\begin{figure}[b]
\center{
\includegraphics[width=\figurewidth]{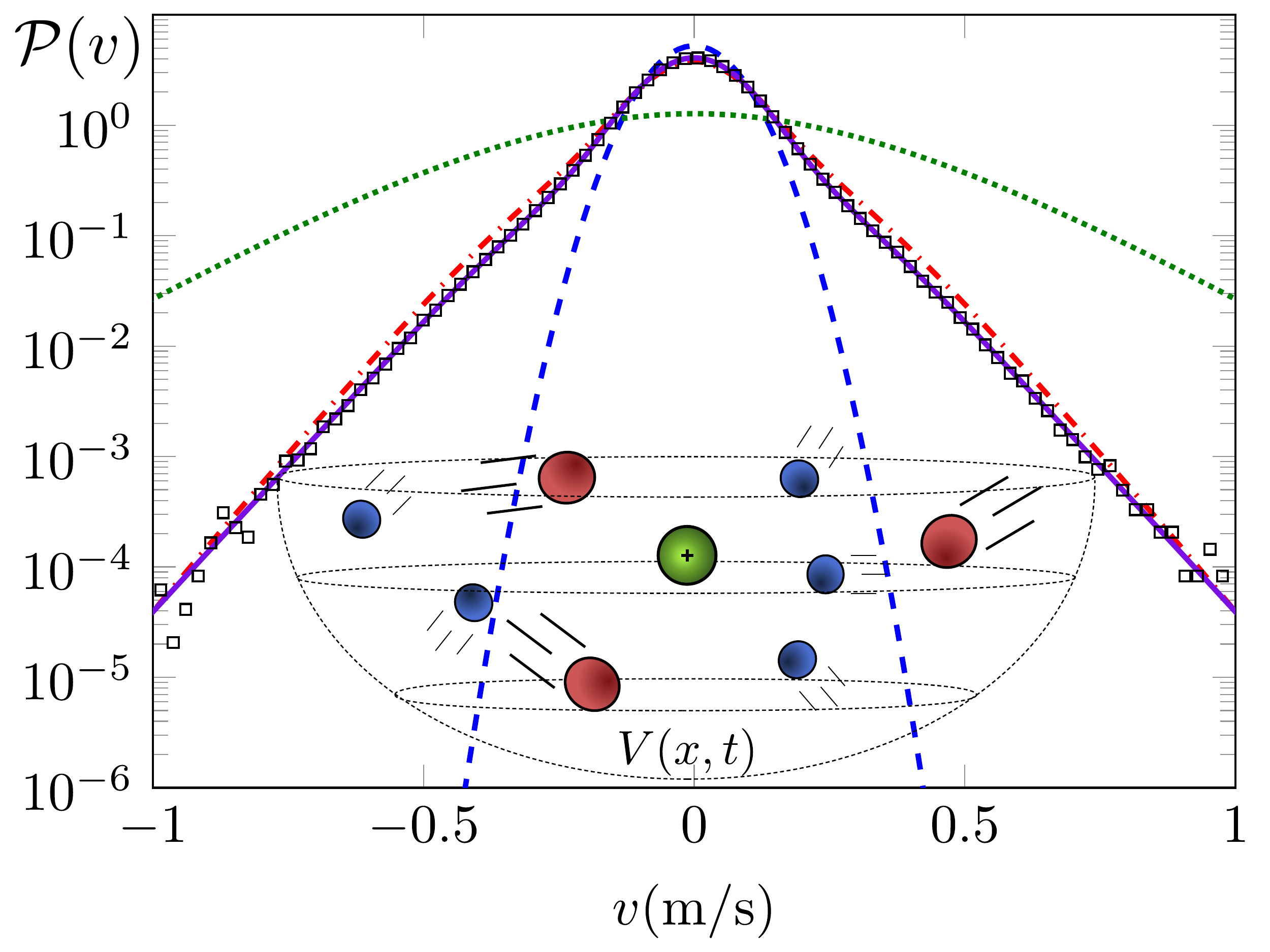}}
\caption{(color online) Steady-state ionic velocity distribution ${\cal P}(v)$ for a two-temperature buffer with 
$T_{\rm c} = 1 (\mbox{m/s})^{2} = 0.01 T_{\rm h}$ with no 
trap.  $m_{\mathrm{c}}/m_{\mathrm{h}} = 20$, $m_{\mathrm{c}}/M = 40/173$.  The distribution is shown for a 
Langevin ion-atom cross section (purple, solid) and a geometric 
cross section (red,dashed-dotted). The power-law velocity tails agree with a Monte Carlo simulation of the velocity transitions (open squares), and 
are distinct from the Gaussian MB distribution of the cold atoms (blue dashed line). 
Time-averaged ${\cal P}(v)$ in a rf trap with $\Omega = 2\pi\mathrm{s}^{-1}$ and a spring constant $k = 100\, s^{-2}$ and using 
the Langevin collision cross section is shown by the (green) dotted line.}
\label{fig:VelocityDistributions}
\end{figure}
The buffer gases constitute large enough thermal reservoirs to maintain their temperatures in spite of the exchange of energy with the ion, and are
decoupled from the external force $F(x,t)$, which acts solely on the ion. 

This master equation was studied by Alkemade and van Kampen~\cite{Alkemade:62}, 
where it was assumed that the  ionic mass $M$ was sufficiently large 
compared to those of the buffer gas atoms that one may expand $W$ for small momentum transfer.  This allowed the replacement of the above 
integro-differential equation with a differential equation containing a small expansion parameter related to the ratio of the buffer gas mass to the tracer particle mass. 
In order to account for hot and heavy buffer gasses we avoid this Kramers-Moyal expansion. 

We solve Eq.~\ref{eq:MasterEquation} analytically (when possible) or numerically using semi-Lagrangian and quadrature methods~\cite{Boyd:01}. 
The solutions are compared to the results of a full three-dimensional molecular dynamics simulation of the trapped ion~\cite{Chen:13}, 
which has been verified against the experimental system. We consider two forcing terms: (i) a simple harmonic trap $F = - k\, x$, 
and (ii) a simple model of the rf trap in one-dimension $F = -k\, x \cos(\Omega t)$.   
\begin{figure}[ht]
\includegraphics[width=\figurewidth]{StaticTrap.pdf}
\includegraphics[width=\figurewidth]{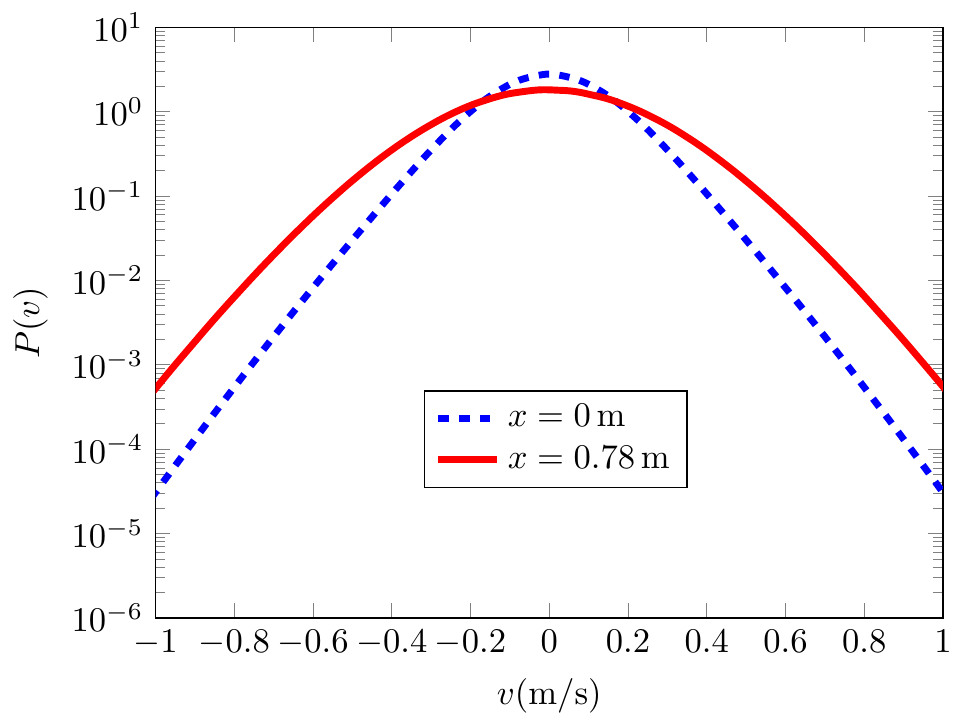}
\caption{(color online) Top: Steady state probability distribution $P(x,v)$ in the same two-temperature buffer gas -- Fig. 1 -- and in a static harmonic potential, 
with  spring constant $k_0=100\,\mathrm{s}^{-2}$. Bottom: Comparison of the velocity distributions at different positions 
(shown by the solid and dashed lines in the top figure), showing position-velocity sorting.}
\label{fig:StaticDynamicTrap}
\end{figure}

In Fig.~\ref{fig:VelocityDistributions} we show the steady-state ionic velocity distributions for various collision cross sections and forcings. These 
are calculated by evolving from an equilibrium distribution at $T_\mathrm{c} = (1/2) v^{2}=1 $ (m/s)$^{2}$. We work in 
units where $k_{\rm B}=1$ and ${\rm amu}=1$.  After 
the hot buffer gas ($T_\mathrm{h} = 100\, T_\mathrm{c}$) is introduced, the initial ionic MB velocity distribution consistent 
with temperature $T_{\rm c}$ (blue, dashed line) broadens. 
After a few mean collision times with the hot buffer gas, it reaches the new steady-state non-MB velocity 
distribution (purple, solid line) having power-law high-velocity tails, which are significantly enhanced as compared to the $T_\mathrm{c}$-MB distribution. 
The nonequilibrium steady-state velocity distribution agrees with Monte Carlo simulations (open squares) of the system.  The appearance 
of the power-law tails occurs in all cases including the rf trap with a Langevin ion-atom cross section (purple, solid line) or in a static trap 
with a simple geometric scattering cross section  (red,dash-dotted line). 

Such power-law tails were also reported~\cite{DeVoe:09} for an
ion interacting with a single-temperature neutral buffer in an rf trap. Multiplicative noise associated with the stochastic (with respect to the rf phase) 
interruption of ionic micromotion alone can account for these~\cite{Chen:14}. Indeed, the high-velocity power-law tails that we 
observe in the {\em absence of micromotion} are strongly enhanced in our model of the rf trap (dotted, green line), so a variety of nonequilibrium forcing methods 
generate this particular feature.

One of the foundational principles of classical statistical mechanics is that the joint probability distribution of the position and momentum degrees of freedom factorizes:  
$P(x,v) = {\cal P}(v; T) {\cal P}(x;T)$ -- e.g., an isothermal gas in a gravitational field has the same MB velocity distribution at each height even though
its density decreases with height. This factorizibility does not survive in the nonequilibrium system of current interest.   
In Fig.~\ref{fig:StaticDynamicTrap} (upper panel)  we plot the ionic steady-state joint probability distribution $P(x,v)$ in a static harmonic trap~\cite{Campbell:07}, 
with spring constant $k_{0} = 100$~s$^{-2}$. In the lower panel, we consider two normalized velocity distributions at different positions: $x_{1}=0$ (blue, dashed) 
\begin{figure}[h]
\includegraphics[width=8.5cm]{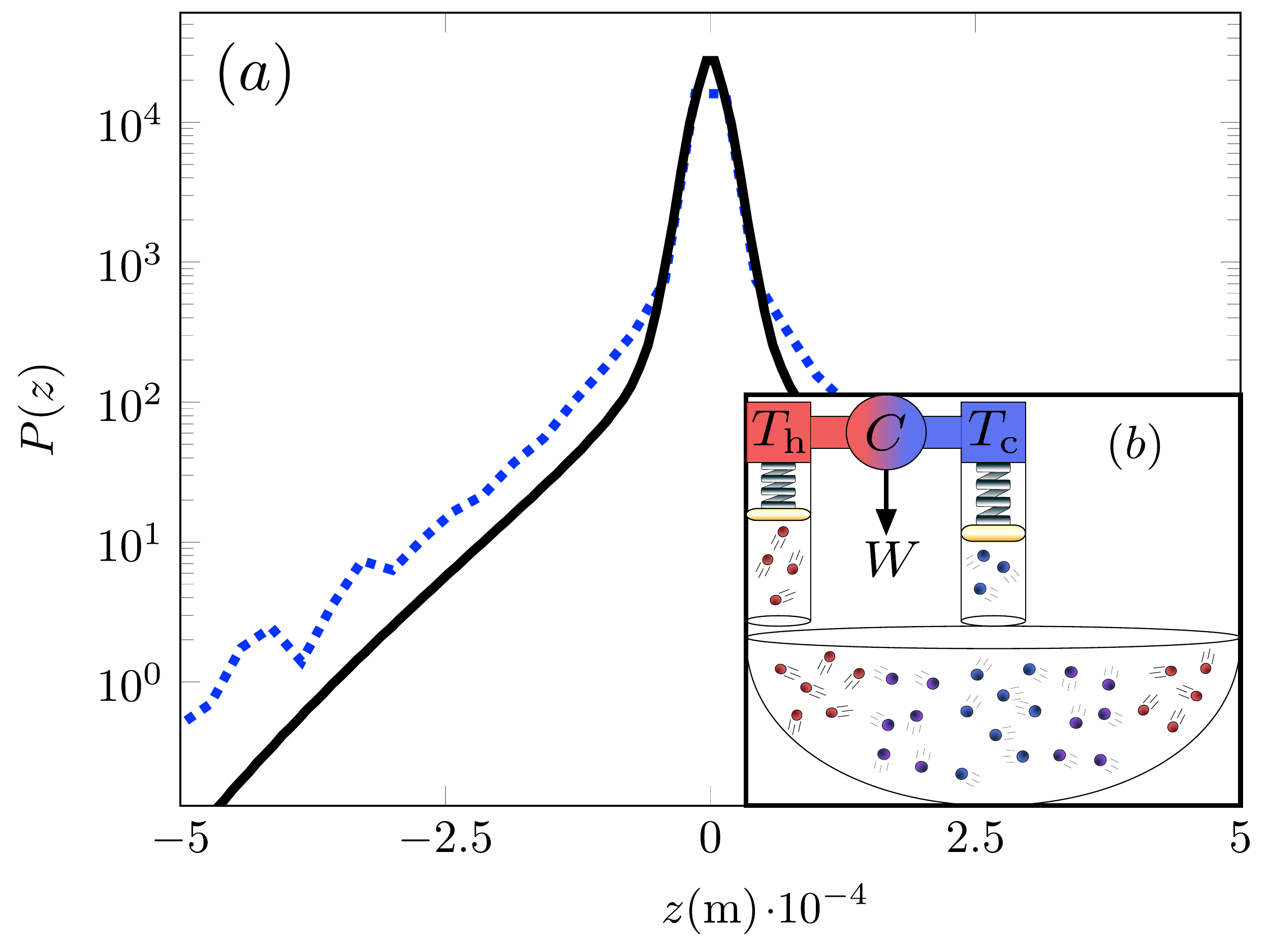}
\caption{(color online)  (a) Ionic vertical position distribution $P(z)$ in an rf trap computed from a 
3D molecular dynamics simulation (blue, dashed) and from the 1D master equation 
(black, solid) using the same simulation parameters for the trap and collision cross section. 1D buffer concentrations were adjusted as a free parameter but the ratio 
$c_\mathrm{h}/c_\mathrm{c} = 10^{-2}$ was fixed by simulation parameters. (b) Schematic Carnot engine used to extract work from the position-velocity sorted state.}
\label{fig:Carnot}
\end{figure}
and $x_{2}=0.78$ m (red, line) in the static trap, and see that the high-velocity states are over represented at higher potential energy (red, line) 
relative to those at zero potential energy at $x_{1}$.  This demonstrates that the joint probability distribution is non-factorizible --  
a feature we term {\em position-velocity sorting.}  Conversely, integrating over all velocities at a given $x$ to 
obtain the spatial distribution, one finds that the potential energy microstates are not populated according a Boltzmann relation  
$\sim \exp[-U(x)]$. We speculate that position-velocity sorting arises from the fact that rare hot atom collisions with the ion typically 
drive it simultaneously to high velocity states and out of the center of the trap, leading to a nonequilibrium correlation 
between fast states and high potential energy ones. This idea is supported by examining the time evolution of the position velocity probability distribution from that of 
a cold ion at the center of the trap to its steady-state form -- see supplemental materials. There one sees that the probability of large displacements from the trap center
occurs coincidently with the appearance of higher probability density of high kinetic energy states, suggesting that one or only a few collisions with hot neutral atoms 
are required to simultaneously propel the ion into high potential energy and high kinetic energy states.

The nonequilibrium ionic spatial probability distribution in an rf trap, shown in Fig.~\ref{fig:Carnot}, is similar to that predicted for the 
static trap  -- see Fig.~2.  In both cases the high potential energy states are overrepresented in the form of power-law tails.  
The one-dimensional calculation (solid, black line) based on  Eq.~1 provides a good fit to the distribution of the ion's $z$ coordinate 
as computed from a molecular dynamics simulation of a three-dimensional rf trap 
with the two-temperature buffer, using experimental realizable parameters.  To compare the one-dimensional theory to 
three dimensional simulation, we 
adjusted the one-dimensional buffer concentrations in the calculation as a single fitting parameter. 

Because of position-velocity sorting, it is, in principle, possible to extract free energy from the system to run a heat engine. For this to happen 
one needs a nonequilibrium system that also breaks a spatial symmetry to generate the directional movement necessary to do 
work on its surroundings. An example is a molecular 
motor~\cite{Prost:97} in which the chemical potential difference between ATP and ADP can generate work only if the motor interacts with 
a directional track (e.g. F-actin) breaking inversion symmetry. We find that, for an ion in the two-temperature buffer, 
{\em both the static and rf harmonic traps} 
provide the necessary symmetry breaking to generate work via {\em position-velocity sorting}.

In Fig.~\ref{fig:Carnot} (b) we show a schematic representation of a Carnot engine (CE) extracting energy out of the trapped ion system to generate work. 
No equilibrium system, such as the thermal reservoirs of the CE, can come into equilibrium with the nonequilibrium steady-state of the ion. Rather, we imagine two thin 
wires coming from the hot and cold thermal reservoirs of the CE allowing energy and momentum exchange with an ensemble of ions at the 
edge and the center of the static trap, respectively. Momentum transfers between the wire and ion are balanced by the wires' 
elastic deformation, represented by springs.  The temperatures of  two reservoirs are adjusted so 
that the net energy transfer vanishes between them and the ensemble of ions.  If these were two equilibrium systems, such a balance 
would imply equal temperatures and pressures, but neither thermodynamic variable is 
meaningful for the ion. Nevertheless, the vanishing net energy flow between the two thermal reservoirs and the ionic ensemble allows us to assign nominal 
temperatures to both of them. When computed this way, the temperature of the hot reservoir in thermal contact with the 
edge of the trap is greater than that of the cold reservoir in thermal contact at the trap center. It is  then possible to generate work in the usual way. 
For the buffer gas parameters used in Fig.~2 (cold reservoir at $x=0\,\mathrm{m}$; hot reservoir at $x=0.39\,\mathrm{m}$) we find a thermodynamic efficiency 
of 0.166.

Another robust feature of equilibrium systems is the relation between their fluctuations and linear response,
known as the fluctuation-dissipation theorem (FDT)~\cite{Chaikin:95}.  In some active matter systems of biological interest, the breakdown of the 
FDT is used as an indicator of nonequilibrium physics associated with endogenous molecular motor activity~\cite{Mizuno:07,MacKintosh:08}. 
Unlike in that system, however, the two-temperature buffer gas provides a
simple and independent control on the nonequilibrium nature of the system -- the temperature of the hot gas. 
As $T_\mathrm{h}$ is reduced to $T_\mathrm{c}$, the system must return to equilibrium. 

The FDT implies a relation between the mean ionic velocity in the presence of a 
static electric field and its diffusion in the absence of one. We test this Einstein relation by applying a constant force $U = - F_{0}\, x$ and 
determining the mean ionic velocity in steady-state. By examining the the ratio of that ionic mean velocity to the 
applied force in the linear response regime, one extracts the ionic mobility $\mu$. 
\begin{figure}[t]
\includegraphics[width=\figurewidth]{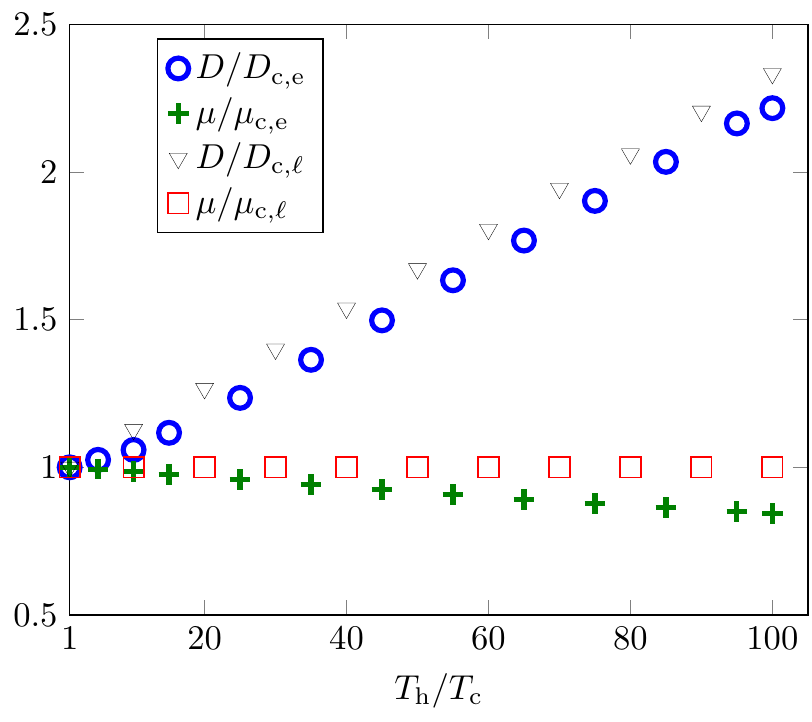}
\caption{(color online) Comparison of the ion diffusivities and mobilities in a two-temperature buffer for a geometric ($\mathrm{e}$) and Langevin 
($\ell$) atom-ion  cross sections. We drive the system from equilibrium by 
controlling $T_{\rm h}/T_{\rm c}$.  Diffusivities and mobilities are normalized by their equilibrium values: $D_{\mathrm{c,i}}$ and $\mu_{\mathrm{c,i}}$ respectively, 
$\mbox{i = e},\ell$.}
\label{fig:FDT}
\end{figure}
Alternatively, by placing the ion at a given initial location and examining the spread of its spatial probability distribution {\em without an applied force},
one obtains the ionic diffusion constant from $D=\lim_{t\to\infty}\langle x^2\rangle/2\,t$. 

Figure~\ref{fig:FDT} shows $\mu$ and $D$ as a function of the buffer gas temperature ratio $T_{\rm h}/T_{\rm c}$, used to control the nonequilibrium steady-state. Both 
$\mu$ and $D$ separately normalized by their equilibrium values, computed by setting the buffer gas temperatures to be equal:  $T_\mathrm{h} \rightarrow T_\mathrm{c}$. 
As the temperature of the hot buffer gas is increased to drive the system from equilibrium, the ionic diffusivity $D$ increases roughly linearly. 
We attribute this to the higher occupation probabilities of high velocity states.  
Though rare, collisions with the hot atoms enhance ionic diffusion relative to its equilibrium value at $T_\mathrm{h} = T_\mathrm{c}$. 
For  the geometric collision cross section $\mu$ decreases since the ion's collision rate with the buffer gas increases with relative velocity. 
For  the Langevin cross section, 
$\mu$ is  constant because the collision rate is independent of the relative velocities of the ion and the buffer atoms.
The Einstein relation breakdown  due to the presence of even a low concentration of a hot buffer gas 
makes the analysis of charge transport in cold plasmas~\cite{Cote:00} more difficult, requiring one to independently consider diffusivity and mobility. 

We have shown that, due to the combination interactions with a hot and cold neutral buffer gas, the ionic joint 
probability distribution for position and velocity in a trap shows a number of striking nonequilibrium features  that can be quantitatively 
controlled by the temperature difference between the two buffers. In  addition, ionic conductivity and diffusion should demonstrate marked 
departures from the FDT. Experimental observation of these effects is, in principle, straightforward.  
A single ion can be prepared at the center of a linear quadrupole trap with weak axial confinement by strong laser cooling. The ion can be released by 
extinguishing the laser cooling, allowing the ion to interact with the buffer gases and thereby explore the system phase space. 
After an evolution time, the 
ion's position can be measured using radial ejection~\cite{Schneider:14,Trippel:09} from the ion trap onto an imaging micro-channel plate (MCP). 
By repeating the experiment, the probability distribution of the ion's position and analogs of the diffusion constant can be measured, 
as well as their dependence on the experimental parameters, such as buffer gas temperature and density, explored. Further, it may be 
possible to use a retarding potential in front of the MCP, or a variation of velocity map ion imaging~\cite{Wester:14}, to measure the ion's kinetic 
energy as a function of trap position to explore position-velocity sorting. It may also be possible to build up the ion position distribution by stroboscopic
laser imaging, as long as the duty cycle is low enough to not significantly affect ion dynamics. Finally, using the same system, the laser cooling can be 
adjusted to prepare the ion with a displacement from the trap minimum. By monitoring the transport of the ion in response to the 
axial trapping force analogs of the ion mobility can be measured.

AJL and CV acknowledge support from NSF-DMR-1309402. EH and KC thank ARO, ARO-MURI, and NSF for partial support.

\bibliography{HotColdRefs}

\end{document}